\documentstyle[preprint,aps]{revtex}
\tightenlines
\input{psfig.sty}
\begin{document}
\preprint{UNDPDK-00-03b}
\title{Bounds on Extra Dimensions from Binary Pulsars}
\author{J.M.~LoSecco}
\address{University of Notre Dame, Notre Dame, Indiana 46556}
\date{\today}
\maketitle
\begin{abstract}
Evidence for gravitational radiation from binary pulsars places
constraints on properties of ``large'' extra dimensions.  The size of these
extra dimensions must exceed about 37 AU for gravitational radiation to
be emitted.\\

Subject headings: string theory --- gravitation --- relativity\\
\\
\end{abstract}
\pacs{PACS numbers: 11.25.Mj, 12.10.-g}

\section{Introduction}
It is widely known that confined systems have a ground state energy that
increases as the confining dimensions are reduced.  The presence of this
ground state energy means that there will be an energy gap between the vacuum,
or zero particle state, and the single particle state.  This gap is the minimum
energy that is required to produce a single quanta.  Long wavelength, low
frequency radiation of light or gravity is a probe of the presence of any
confining dimensions.  A bound on these dimensions follows from the minimum
energy needed to create a quanta.  Quanta with energies less than the gap
energy can not be excited.  Dissipationless motion will occur for orbital
periods with $T > h / E_{gap}$.

\section{Particle in a Box}
A particle rigidly confined to a region of size L has a ground state
momentum associated with this confinement of $p = \frac{h}{2L}$.
A component of momentum, of this order or greater is associated with each of
the confining dimensions.  Even with only one confining dimension, this puts
a lower bound of $E > \frac{hc}{2L}$ for massless particles.  A contribution
to the energy is associated with each of the confining dimensions.

It has been suggested that the extra dimensions associated with string theory
might be compactified on a distance scale large compared to the Planck
length,\cite{larged}.
For a single extra dimension of size $L$ the lowest momentum associated with
confinement in this dimension is a standing wave of length $2L$ with a
momentum of $\frac{h}{2L}$ and an energy $\frac{hc}{2L}$.  If the confinement
is not ``rigid'' lower energies are possible.  If the confining region is not
``flat'' higher energies are needed.  A minimum energy means that radiation
of lower energy (long wavelength) is suppressed.

\section{Gravitational Radiation from a Binary Pulsar}
The inference of gravitational radiation from the rotational period of a
binary pulsar places a lower limit on the size of any extra dimensions that are
accessible to forces in our 4 dimensional world.  In particular the long
wavelength gravitational radiation probes extra dimensions accessible to
gravity.  Similar bounds can be established for electromagnetic radiation,
but photons may not be sensitive to these extra dimensions.

The period of gravitational radiation has been reported \cite{bpuls} to be
about $T=$36351.70270(3) seconds for PSR 1534+2.  The quanta associated with
this radiation have energies of about $1.8228 \times 10^{-38}$ Joules, or
$1.1377 \times 10^{-19}$ eV.

\[
E = pc = \frac{ h c }{\lambda}
\]
\[
\lambda < 2 L
\]
where L is the confining dimension.
\[
L > \frac{h c}{2 E} = \frac{c T}{2}
\]

Which yields a a bound of $L > 5.489 \times 10^{12}$ meters.  This distance is
about 37 times 1 AU, the distance from the Earth to the Sun.  To avoid
suppression of the gravitational radiation, the additional
dimensions must be large compared to this scale.

To evade this bound would require profound modifications to our concepts
of space-time or quantum mechanics.  Alternatively, one might argue that the
evidence for gravitational radiation is indirect.  The pulsar orbital period
is decreasing.  This change can be explained by gravitational radiation,
with a period equal to the orbital period.  General relativity and the
observations seem to agree but the gravitational radiation itself has not
been detected.

\section{Acknowledgements}
I would like to thank G.~Domokos and J.~Hewett for useful comments.
N.~Uraltsev and C.~Kolda have kindly commented on an earlier version
of this manuscript.
Correspondence with Bogdan Dobrescu is gratefully acknowledged.

\end{document}